\documentclass[twocolumn,tighten]{aastex701}

\usepackage{xspace}
\usepackage[figuresright]{rotating}
\usepackage{enumitem}
\usepackage{comment}
\usepackage{amsmath}

\newcommand{\odin}{One-hundred-deg$^2$ DECam Imaging in Narrowbands\xspace}

\newcommand{\lya}    {Ly$\alpha$\xspace}

\newcommand{\msun}   {$M_{\odot}$\xspace}

\newcommand{\hof}     {$f_\mathrm{LAB}$\xspace}

\newcommand{\unitcgssb}   {erg\,s$^{-1}$\,cm$^{-2}$\,arcsec$^{-2}$\xspace}

\newcommand{\unitcgslum}  {erg\,s$^{-1}$\xspace}

\newcommand{\kms}         {km\,s$^{-1}$\xspace}

\newcommand{\sqarcsec}    {arcsec$^{2}$\xspace}
\newcommand{\sqarcmin}    {arcmin$^{2}$\xspace}
\newcommand{\sqdeg}    {deg$^{2}$\xspace}

\definecolor{forestgreen}{rgb}{0.13, 0.55, 0.13}

\definecolor{brown}{rgb}{0.52, 0.33, 0.14}

\shorttitle{Clustering Analysis For Lyman Alpha Blobs}
\shortauthors{Moon et al.}
\graphicspath{{./}{figures/}}

\begin{document}

\title{ODIN: Clustering Properties of \lya Blobs at $z$ $\sim$ 2.4 and 3.1}

\author[0009-0008-4022-3870]{Byeongha Moon}
\affiliation{Korea Astronomy and Space Science Institute, 776 Daedeokdae-ro, Yuseong-gu, Daejeon 34055, Korea}
\affiliation{Korea National University of Science and Technology, Daejeon 34113, Korea}
\email[show]{bhmoon@kasi.re.kr}

\author[0000-0003-3078-2763]{Yujin Yang}
\affiliation{Korea Astronomy and Space Science Institute, 776 Daedeokdae-ro, Yuseong-gu, Daejeon 34055, Korea}
\affiliation{Korea National University of Science and Technology, Daejeon 34113, Korea}
\email{yyang@kasi.re.kr}

\author[0000-0003-1530-8713]{Eric Gawiser}
\affiliation{Department of Physics and Astronomy, Rutgers, the State University of New Jersey, Piscataway, NJ 08854, USA}
\email{gawiser@physics.rutgers.edu}

\author[0000-0003-3004-9596]{Kyoung-Soo Lee}
\affiliation{Department of Physics and Astronomy, Purdue University, 525 Northwestern Avenue, West Lafayette, IN 47907, USA}
\email{soolee@purdue.edu}

\author[0000-0003-2986-8594]{Danisbel Herrera}
\affiliation{Department of Physics and Astronomy, Rutgers, the State University of New Jersey, Piscataway, NJ 08854, USA}
\email{dh974@rutgers.edu}

\author[0000-0002-9176-7252]{Vandana Ramakrishnan}
\affiliation{Department of Physics and Astronomy, Purdue University, 525 Northwestern Avenue, West Lafayette, IN 47907, USA}
\email{ramakr18@purdue.edu}

\author[0000-0001-9850-9419]{Nelson Padilla}
\affiliation{Instituto de Astronomía Teórica y Experimental, CONICET-Universidad Nacional de Córdoba, Laprida 854, X5000BGR, Córdoba, Argentina}
\email{n.d.padilla@gmail.com}
\author[0000-0002-9811-2443]{Nicole M. Firestone}
\affiliation{Department of Physics and Astronomy, Rutgers, the State University of New Jersey, Piscataway, NJ 08854, USA}
\email{nmf82@physics.rutgers.edu}

\author[0009-0002-3931-6697]{Seongjae Kim}
\affiliation{Korea Astronomy and Space Science Institute, 776 Daedeokdae-ro, Yuseong-gu, Daejeon 34055, Korea}
\affiliation{Korea National University of Science and Technology, Daejeon 34113, Korea}
\email{seongjkim@kasi.re.kr}

\author[0000-0002-1328-0211]{Robin Ciardullo}
\affiliation{Department of Astronomy \& Astrophysics, The Pennsylvania
State University, University Park, PA 16802, USA}
\affiliation{Institute for Gravitation and the Cosmos, The Pennsylvania
State University, University Park, PA 16802, USA}
\email{rbc3@psu.edu}

\author[0000-0001-6842-2371]{Caryl Gronwall}
\affiliation{Department of Astronomy \& Astrophysics, The Pennsylvania
State University, University Park, PA 16802, USA}
\affiliation{Institute for Gravitation and the Cosmos, The Pennsylvania
State University, University Park, PA 16802, USA}
\email{caryl@astro.psu.edu}

\author[0000-0002-4902-0075]{Lucia Guaita}
\affiliation{Universidad Andres Bello, Facultad de Ciencias Exactas, Departamento de Fisica y Astronomia, Instituto de Astrofisica, Fernandez Concha 700, Las Condes, Santiago RM, Chile}
\affiliation{Millennium Nucleus for Galaxies (MINGAL)}
\email{lucia.guaita@gmail.com}

\author[0000-0003-3428-7612]{Ho Seong Hwang}
\affiliation{Astronomy Program, Department of Physics and Astronomy, Seoul National University, 1 Gwanak-ro, Gwanak-gu, Seoul 08826, Republic of Korea}
\affiliation{SNU Astronomy Research Center, Seoul National University, 1 Gwanak-ro, Gwanak-gu, Seoul 08826, Republic of Korea}
\email{hwang.ho.seong@gmail.com}

\author[0009-0003-9748-4194]{Sang Hyeok Im}
\affiliation{Astronomy Program, Department of Physics and Astronomy, Seoul National University, 1 Gwanak-ro, Gwanak-gu, Seoul 08826, Republic of Korea}
\affiliation{Korea Institute for Advanced Study, 85 Hoegi-ro, Dongdaemun-gu, Seoul 02455, Republic of Korea}
\email{sanghyeok.im97@gmail.com}

\author[0000-0002-2770-808X]{Woong-Seob Jeong}
\affiliation{Korea Astronomy and Space Science Institute, 776 Daedeokdae-ro, Yuseong-gu, Daejeon 34055, Korea}
\affiliation{Korea National University of Science and Technology, Daejeon 34113, Korea}
\email{jeongws@kasi.re.kr}

\author[0000-0001-6270-3527]{Ankit Kumar}
\affiliation{Universidad Andres Bello, Facultad de Ciencias Exactas, Departamento de Fisica y Astronomia, Instituto de Astrofisica, Fernandez Concha 700, Las Condes, Santiago RM, Chile}
\email{ankit4physics@gmail.com}

\author[0000-0002-6810-1778]{Jaehyun Lee}
\affiliation{Korea Astronomy and Space Science Institute, 776 Daedeokdae-ro, Yuseong-gu, Daejeon 34055, Korea}
\email{jaehyun@kasi.re.kr}

\author[0000-0001-5342-8906]{Seong-Kook Lee}
\affiliation{Astronomy Program, Department of Physics and Astronomy, Seoul National University, 1 Gwanak-ro, Gwanak-gu, Seoul 08826, Republic of Korea}
\affiliation{SNU Astronomy Research Center, Seoul National University, 1 Gwanak-ro, Gwanak-gu, Seoul 08826, Republic of Korea}
\email{s.joshualee@gmail.com}

\author[0000-0002-7356-0629]{Julie B. Nantais}
\affiliation{Facultad de Ciencias Exactas, Departamento de Física y Astronomía, Instituto de Astrofísica, Universidad Andrés Bello, Fernández Concha 700, Edificio C-1, Piso 3, Las Condes, Santiago, Chile}
\email{julie.nantais@unab.cl}

\author[0000-0001-9521-6397]{Changbom Park}
\affiliation{Korea Institute for Advanced Study, 85 Hoegi-ro, Dongdaemun-gu, Seoul 02455, Republic of Korea}
\email{cbp@kias.re.kr}

\author[0000-0002-4362-4070]{Hyunmi Song}
\affiliation{Department of Astronomy and Space Science, Chungnam National University, 99 Daehak-ro, Yuseong-gu, Daejeon, 34134, Republic of Korea}
\email{hmsong@cnu.ac.kr}



\begin{abstract}
Spatially extended \lya nebulae, known as \lya blobs (LABs), are a rare population at $z > 2$ that are thought to trace proto-groups or the progenitors of massive galaxies in the present-day universe. However, their dark matter halo properties (e.g.,  halo mass) are still uncertain due to their rarity and strong field-to-field variation.
The \odin (ODIN) survey has discovered 103 and 112 LABs in the extended ($\sim$9~\sqdeg) COSMOS field at $z\sim2.4$ and 3.1, respectively, enabling estimation of their bias and host halo masses through clustering analysis.
We measure the angular auto-correlation functions (ACFs) of LABs and derive galaxy bias factors of
$b$ = $4.0\pm0.8$ and $3.8\pm0.7$, corresponding to minimum halo masses of $2.8^{+3.0}_{-1.8}$ and $0.7^{+0.8}_{-0.5}\times10^{12}$\;\msun and median halo masses of $4.2^{+3.8}_{-2.5}$ and $1.1^{+1.1}_{-0.7}\times10^{12}$\;\msun at $z\sim2.4$ and 3.1, respectively.
LABs occupy $\sim$11$^{+39}_{-8}$\% and $\sim$3$^{+9}_{-2}$\% of all dark matter halos above these minimum halo masses.
These findings suggest that LABs inhabit massive dark matter halos, likely tracing proto-group environments that evolve into present-day massive halos ($\sim$10$^{13}$\;\msun), where massive elliptical galaxies or galaxy groups reside, by $z=0$.
\end{abstract}



\section{Introduction} \label{sec:intro}

\begin{figure*}[th]
\centering
\includegraphics[width=0.95\textwidth]{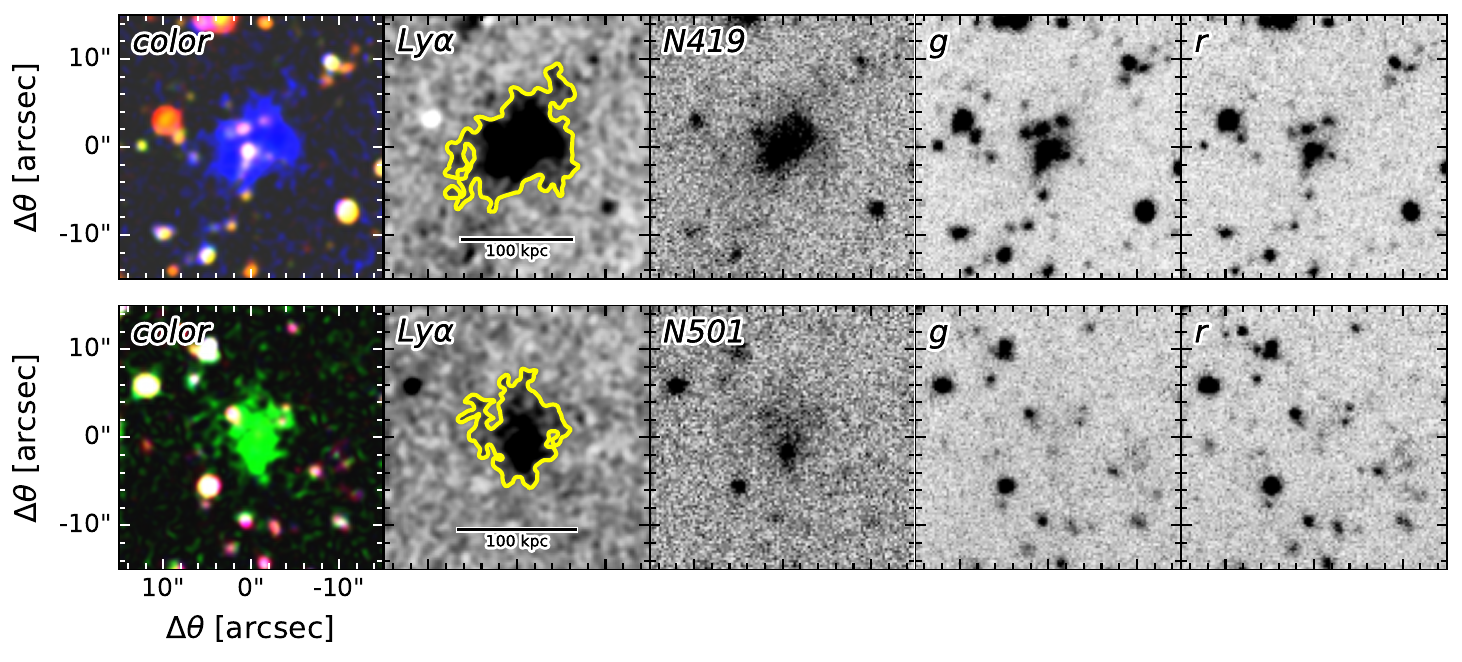}
\caption{
Postage-stamp images ($30\times30$ \sqarcsec) of selected LABs in this study.
In each panel, the yellow contours indicate the surface brightness detection thresholds.
\textbf{Top:} A LAB at $z\sim2.4$, shown in a composite of Subaru/HSC $r$ (red), Subaru/HSC $g$ (green), and ODIN $N419$ (blue) images.
\textbf{Bottom:} A LAB at $z\sim3.1$, shown in a composite of Subaru/HSC $r$ (red), ODIN $N501$ (green), and Subaru/HSC $g$ (blue) images.
}
\label{fig:fig2.LAB_sample.pdf}
\end{figure*}

\lya blobs (LABs) broadly refer to spatially extended ($\gtrsim$\,50\,kpc) \lya emission nebulae, often associated with multiple embedded galaxies, discovered at high redshifts \citep[$z>2;$][]{Keel99, Steidel00, Matsuda04, Yang09, Yang10, Shibuya18, kikuta19, Moon26a}.
They are often found in the outskirts of overdense regions \citep{Matsuda11, Badescu17, Ortiz26}, near and in filamentary structures traced by \lya emitters \citep[LAEs;][]{Umehata19, Ramakrishnan23}, thus showing a strong association with large-scale structure. 
LABs are known to host a diverse mass range of galaxies there in, such as LAEs, Lyman break galaxies (LBGs), star-forming galaxies, active galactic nuclei (AGNs), submillimeter galaxies (SMGs), and QSOs \citep{Matsuda04, Dey05, Geach09, Yang11, Yang14a, Yang14b, Yamada12, Alexander16, Borisova16, Ao17, Kato18, Fabrizio18a, Vale23, Mingyu24}. 
The number of member galaxies within each LAB also varies widely.
Some LABs appear to trace proto-groups, embedding multiple galaxies and/or being surrounded by galaxies \citep{Prescott12a, Umehata21, Daddi20, Daddi22}.
As such, LABs may provide valuable laboratories for studying the formation of massive galaxies and the hierarchical assembly of large-scale structures.
However, the underlying halo properties of LABs are not yet well understood, particularly the dark matter halo mass, which provides critical insights into galaxy formation and evolution.

The simplest tool to link LABs to dark matter halos is clustering analysis via the measurement of the two-point correlation function.
The theoretical framework underlying such clustering analysis, based on the Press-Schechter theory and cosmological $N$-body simulations, have shown that galaxies are biased tracers of the dark matter distribution, with their properties determined by the halos in which they form \citep[e.g.,][]{Press74, White78, Kaiser84, Bardeen86, Mo96, Somerville01, Sheth01}.
Additionally, the Halo occupation distribution \citep[HOD;][]{Berlind02} framework provides a statistical prescription for the halo-galaxy connection by quantifying the probability that a dark matter halo of mass $M$ hosts $N$ galaxies, $P(N|M)$. This approach provides insights into the underlying dark matter halo properties, including the galaxy abundance \citep[e.g.,][]{Soo06, Zehavi11, Geach12, Hong19}.

Clustering analyses have been widely used in studies of high-redshift galaxy populations and have provided key insights  into their evolution. 
For example, LAEs at $z>2$ have been shown to have low galaxy bias ($b\sim2$), indicating that they reside in $10^{11}$\;\msun halos and thus evolve into $\sim$$L^\star$ galaxies by $z=0$ \citep{Gawiser07, Kovac07, Guaita10, Ouchi10, Kusakabe18, Hong19, White24, Umeda25, Herrera25}. 
Simple HOD models further suggest that only 1\% of the dark matter halos above a minimum halo mass of $\sim$10$^9$\;\msun host LAEs at $z\sim6$ \citep{Ouchi18}, increasing to 3--7\% for LAEs at $z\sim2.4$, 3.1 and 4.5 \citep{Herrera25}.
However, the application of clustering analysis to LABs has been limited by their rarity and strong field-to-field variations in number density, arising from the small survey areas explored to date \citep[e.g.,][]{Yang09, Yang10}, although see also \citealt{Zhang25}.

Clustering analysis of LABs within a single contiguous and large field has now become possible with the \odin survey \citep[ODIN;][]{Soo23}. ODIN is a wide-field narrowband imaging survey that covers seven extragalactic fields (a total of $\sim$91\;\sqdeg) with three filters, $N419$, $N501$, and $N673$, targeting \lya-emitting objects at $z\sim2.4$, 3.1, and 4.5.
From the 9~\sqdeg\ extended COSMOS, the ODIN survey has successfully discovered $\sim$6,000 LAEs, $\sim$100 LABs, and $\sim$40 proto-clusters  at $z \sim 3.1$ \citep{Ramakrishnan23, Ramakrishnan24, Ramakrishnan25, Ramakrishnan26, Firestone24, Moon26a}.

In this work, we report the clustering properties of ODIN LABs at $z\sim2.4$ and $3.1$ in the 9~\sqdeg extended COSMOS field.
The paper is organized as follows. 
In Section~\ref{sec:data}, we summarize the ODIN LAB samples.
In Section~\ref{sec:result}, we present the results of our clustering analysis, including the galaxy bias, correlation length, and halo mass of LABs. 
In Section~\ref{sec:discussion}, we discuss the halo occupation fraction of LABs and compare our findings with previous results.
Section~\ref{sec:summary} provides a summary and conclusion.
Throughout our analysis, we assume a flat $\Lambda$CDM cosmology, with $\Omega_{\Lambda}=0.7$, $\Omega_{m}=0.3$, $n_s=0.95$, and $H_0$ = 100$h$\,\kms\,Mpc$^{-1}$ with $h=0.7$.
All distance scales are given in comoving units unless stated otherwise.

\section{LAB Sample} \label{sec:data}
The ODIN survey targets seven fields (COSMOS, Deep2-3, SHELA, XMM-LSS, CDF-S, EDF-S, and ELAIS-S1) with three narrowband filters that have a central wavelength of 4193\;\AA, 5014\;\AA, and 6750\;\AA\ with FWHM of 75\;\AA, 76\;\AA, and 100\;\AA\ corresponding to the line-of-sight thickness of 76, 57 and 51~Mpc \citep{Soo23}.  
The filters are designed to be sensitive to redshifted \lya emission at $z=2.45\pm0.03$, $3.12\pm0.03$ and $4.55\pm0.04$.
In this study, we use the COSMOS LAB samples at $z\sim2.4$ and 3.1, for which the largest numbers of LABs have been discovered within a single contiguous  9~\sqdeg field.

Here, we briefly introduce the LAB samples. We refer readers to \cite{Moon26a} for details of the LAB selection methods adopted for the entire ODIN survey. 
ODIN LABs are defined as sources that satisfy the following criteria: (1) measured \lya isophotal areas larger than  20~\sqarcsec, (2) rest-frame \lya equivalent widths greater than 20~\AA, and (3) sizes larger than those of point sources at a given \lya luminosity.
Figure~\ref{fig:fig2.LAB_sample.pdf} shows examples of LABs at $z\sim2.4$ and 3.1.
By performing isophotal photometry with detection thresholds of $5.2\times10^{-18}$ and $3.3\times10^{-18}$~\unitcgssb at $z \sim 2.4$ and $z \sim 3.1$, respectively, we discover 103 and 112 LABs within survey volumes of $6.7\times10^6$ Mpc$^3$ and $6.5\times10^6$ Mpc$^3$.
The survey volumes are calculated from the effective survey area, assuming the FWHM of the narrowband filters as the line-of-sight thickness.
The detection thresholds are defined as 1.5$\sigma_{\rm SB_1}$ at each redshift, where $\sigma_{\rm SB_1}$ is the surface brightness limit per 1~\sqarcsec aperture within each field.
Given the survey area and the line-of-sight thickness sampled by the ODIN narrowband filters, we obtain LAB number densities of $n$ = $1.6\times10^{-5}$ and $1.7\times10^{-5}$~Mpc$^{-3}$ at $z\sim2.4$ and 3.1, respectively.

\begin{figure*}[t]
\centering
\includegraphics[width=0.9\textwidth]{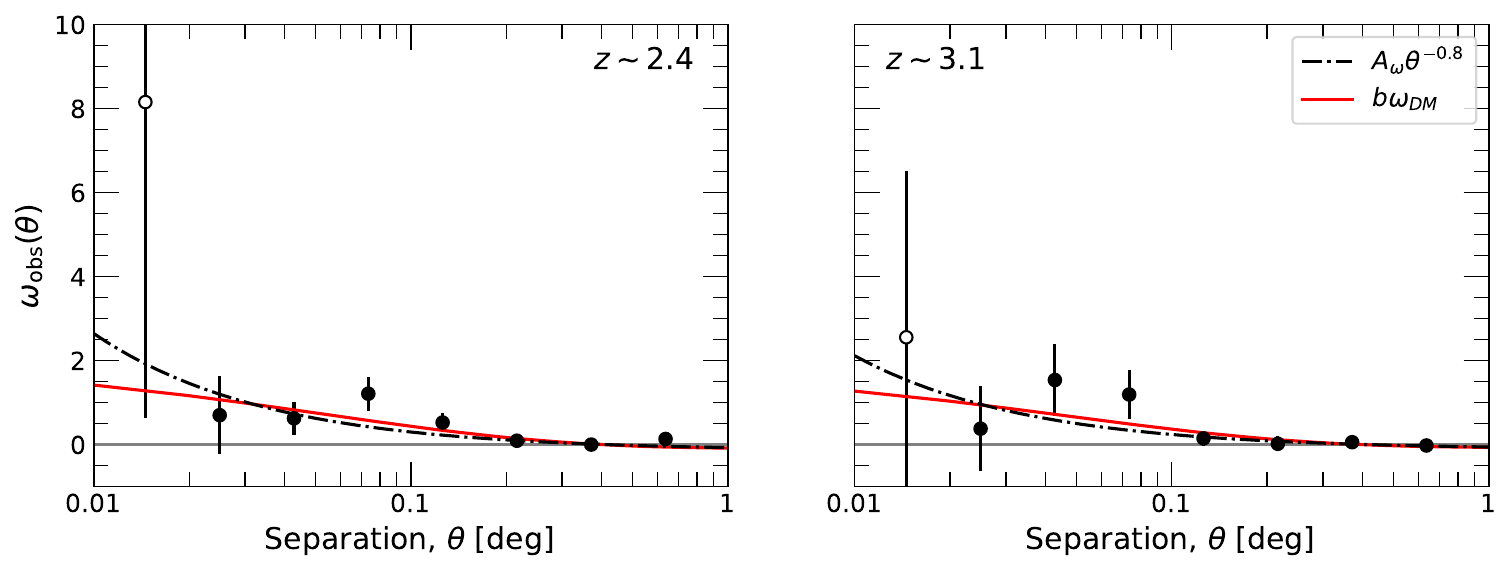}
\caption{Observed ACFs of LABs at $z\sim2.4$ and $3.1$ [$\omega_\mathrm{obs}=(\omega-\mathrm{IC}) / (1+\mathrm{IC})$]. Open markers are not used for fitting the correlation functions because they could be the one-halo term. 
Black dash-dot lines represent the best fit power-law function (Equation~\ref{eq:ACF_fit_func}, $\beta=0.8$), while red solid lines are the best fit with Equation~\ref{eq:linear_bias}. 
The best-fit parameters are summarized in Table~\ref{tab:acf_result}.
}
\label{fig:acf_fit}
\end{figure*}

\section{Analysis \label{sec:result}}
\subsection{Angular Auto--Correlation Function of LABs \label{sec:ACF}}
We calculate the two-point angular correlation function (ACF) of LABs using the \citet{Landy_Szalay93} estimator:
\begin{equation}
    \omega_{\rm obs}(\theta) = \frac{D\!D(\theta)-2D\!R(\theta)+R\!R(\theta)}{R\!R(\theta)},
\end{equation}
where $D\!D(\theta)$, $D\!R(\theta)$, and $R\!R(\theta)$ are the numbers of LAB--LAB pairs, LAB--random pairs, and random--random pairs within the angular annulus $\theta\pm\Delta\theta/2$ in logarithmic space. 
We set the angular separation bins from $\theta=40$\arcsec\ to $\theta = 3000$\arcsec\ with a bin size of $\Delta \log{\theta} = 0.098$. The first bin starts at 40\arcsec\ ($\sim$1.2 Mpc at $z\sim2.4$ and 3.1), corresponding to the minimum separation between LABs. Each pair count is normalized by the total number of pairs.
We construct a random galaxy catalog containing 100 times as many objects as the observed LABs. 
The probability of placing a random galaxy at a given position is weighted by the inverse variance maps of the narrowband image to account for surface brightness fluctuations \citep[e.g.,][]{Morrison15}. 
Random sources within the masked regions used for LAB selection---such as those around bright stars and Galactic cirrus---are excluded \citep{Moon26a}.

The uncertainty of the ACF is estimated using the jackknife method. We divide the ODIN survey area into $3 \times 3$ squares ($1 \times 1$ \sqdeg\, each covering the maximum angular separation, $\theta=3000$\arcsec) and compute the ACFs using eight subregions, leaving out one of the nine subregions in each iteration. We obtain the uncertainty from the covariance matrix from $N=9$ jackknife resamplings \citep{Norberg09}.

Because the finite survey area limits the sample size,
our correlation amplitude will be underestimated.  To avoid this, we apply an integral constraint \citep[IC, ][]{Roche99}:
\begin{equation} \label{eq:IC}
    \mathrm{IC} ~=~ \frac{\sum_{i}{ [R\!R(\theta_i)~\omega(\theta_i)]} }{\sum_{i}{R\!R(\theta_i)}}   
\end{equation}
where $i$ refers $i$-th angular bin.
If the intrinsic ACF, $\omega(\theta)$, can be approximated by a power law index $\beta$: 
\begin{eqnarray} \label{eq:ACF_fit_func}
    \omega(\theta) &~=~& A_\omega \theta^{-\beta}, \nonumber \\
    \omega_\mathrm{obs}(\theta) &~=~& \frac{\omega(\theta) - \mathrm{IC}}{1+\mathrm{IC}}
\end{eqnarray}
the observed ACF, $\omega_\mathrm{obs}(\theta)$, can be fit with Equation~\ref{eq:ACF_fit_func}, using $A_\omega$ and $\beta$ as free parameters.

Figure~\ref{fig:acf_fit} shows the observed ACFs of LABs at $z \sim 2.4$ and $z \sim 3.1$. The red solid lines indicate the best-fit power-law function with the index $\beta$ fixed to 0.8 \citep[e.g.,][]{Peebles80, Kovac07}. 
Because the first bins ($40\arcsec \leq \theta < 70\arcsec$, corresponding to 1--2 Mpc at both redshifts) have large uncertainties and their elevated amplitudes suggest that we may be detecting one-halo clustering, we exclude them from the fitting. 
The one-halo term could arise from LABs residing within the same dark matter halo of a proto-cluster, whose median effective radius is approximately 4--5 Mpc at $z\sim2.4$ and 3.1  \citep{Ramakrishnan24, Ramakrishnan25}.
A more detailed HOD analysis, including the one-halo term, will be explored in future work.
The results are summarized in Table~\ref{tab:acf_result}. We note that the results are not significantly affected by  the first or even the second bin, nor by allowing the parameter $\beta$ to vary freely.

\subsubsection{Correlation Length and Bias} \label{subsec:bias}
From the ACF measurements, we compute the correlation length, $r_0$,  using Limber's equation \citep{Limber1953}:
\begin{eqnarray} \label{eq:corr_length}
    A_\omega &~=~& \frac{C_\gamma H_0 r_0^\gamma}{c} \int_{0}^{\infty}{(dN/dz)^2 \chi(z)^{1-\gamma} E(z) dz} \\
    C_\gamma &~=~& \frac{\Gamma(1/2) \Gamma[(\gamma-1)/2]}{\Gamma(\gamma/2)} \nonumber \\
    E(z) &~=~& \left [\Omega_{\Lambda} + \Omega_{m} (1+z)^3 \right ] ^{1/2}, \nonumber 
\end{eqnarray}
where $\gamma=1+\beta$ and $\Gamma$ is the Gamma function.
The $dN/dz$ denotes the redshift distribution of LABs, normalized such that the integral over redshift equals unity.
We use the redshift distribution of ODIN LAEs from DESI observations \citep{White24, Pinarski26}, assuming that ODIN LAEs and LABs share a similar redshift distribution, given that both are identified using the same narrowband filters and equivalent width threshold. Here, $\chi(z)$ is a transverse comoving distance. 

From Equation~\ref{eq:corr_length}, we estimate the correlation length of LABs: $r_0=6.5\pm1.0$~$h^{-1}$\;Mpc at $z\sim2.4$ and $r_0=5.2\pm1.4$~$h^{-1}$\;Mpc at $z\sim3.1$. These values are larger than those of LAEs \citep[$r_0\sim$ 3--4 $h^{-1}$\;Mpc,][]{Gawiser07, Guaita10,  Kusakabe18, White24, Umeda25}, but smaller than those of proto-clusters \citep[$r_0\sim$ 12--35 $h^{-1}$\;Mpc,][]{Toshikawa18, Ramakrishnan25} at similar redshifts.

Although the Limber approximation may be inaccurate for narrowband observations at high redshifts \citep[e.g.,][]{Simon07, Ouchi18}, we adopt it here to facilitate comparison with other clustering studies that also use the Limber approximation for narrowband data \citep[e.g.,][]{Herrera25, Ramakrishnan25, Zhang25}. We further note that we also perform a numerical integration of the exact equation following \cite{Simon07} and find results consistent with those obtained using the Limber approximation, similar to the agreement reported by \cite{Khostovan18} for their sample.

Galaxy bias describes the relationship between the spatial distribution of galaxies and the underlying dark matter, quantifying how strongly galaxies are clustered relative to the matter. Here, we measure the bias of LABs in two ways. The first is via the ratio between the source distribution and the underlying matter distribution on a scale of 8 $h^{-1}$\;Mpc, 
\begin{eqnarray} \label{eq:bias_sigma}
    b_{\sigma_8} &=& \frac{\sigma_\mathrm{8,LAB}}{\sigma_8(z)}, \\
    \sigma_\mathrm{8,LAB} &=& \sqrt{J_2(\gamma) \biggl( \frac{r_0} {8~h^{-1}~\mathrm{Mpc}} \biggr)^\gamma}, \nonumber \\
    \sigma_8(z) &=& \sigma_8(0) D(z), \nonumber 
\end{eqnarray}
where $J_2=72/[2^\gamma(3-\gamma)(4-\gamma)(6-\gamma)]$ and $\gamma=1+\beta$. $D(z)$ is the linear growth factor of density fluctuations.
By adopting $\sigma_8(0)=0.8$, we obtain bias values of $b$ = $3.8\pm0.5$ and $3.7\pm0.9$ for $z\sim2.4$ and $3.1$, respectively.

Second, the galaxy bias can be determined by the ratio between galaxy and matter correlation functions:
\begin{equation} 
\label{eq:linear_bias}
    {b^2_{\omega}}=\frac{\omega(\theta)}{\omega_{m}(\theta)}.
\end{equation}
For the matter correlation function, $\omega_{m}(\theta)$, we use Equation~A5 in \cite{Myers07} derived from the Limber equation:
\begin{eqnarray}
    && \omega_m(\theta) = \nonumber \\
    && \frac{H_0 \pi}{c}\!\!\int_0^\infty\!\!\int_0^\infty\!\!\frac{\Delta^2(k,z)}{k^2}J_0[k\theta\chi(z)]\biggl(\frac{dN}{dz}\biggr)^2 E(z) dk dz. \nonumber
\end{eqnarray}
The $\Delta^2(k,z)$ is the dimensionless matter power spectrum, defined as $k^3 P(k,z) / (2\pi^2)$, and $P(k,z)$ is the linear dark matter perturbation spectrum computed using the Core Cosmology Library \citep[CCL;][]{Chisari19}.
By combining this bias relation with Equation~\ref{eq:ACF_fit_func}, we directly estimate the galaxy bias from the observed auto-correlation functions without the need to emphasize the particular scale of 8 $h^{-1}$\;Mpc.  
We assume the same $dN/dz$ as in Equation~\ref{eq:corr_length}, the redshift distribution of ODIN LAEs. 

\begin{deluxetable}{c cc l cc}
\centering
\tablecaption{Summary of ACF clustering properties of ODIN LABs \label{tab:acf_result}}
\tablewidth{0pt}
\tablehead{
\colhead{} &
\multicolumn{2}{c}{Bias} &
\colhead{} &
\multicolumn{2}{c}{Power-law fit, $\beta=0.8$} \\
\cline{2-3}
\cline{5-6}
\colhead{$z$} & 
\colhead{$b_{\omega_\mathrm{DM}}$\tablenotemark{1}} & 
\colhead{$b_{\sigma_8}$\tablenotemark{2}} &
\colhead{} &
\colhead{$A_\omega$} & 
\colhead{$r_0$ [$h^{-1}$~Mpc]} 
}
\startdata
2.4 & \textbf{4.0$\pm$0.8} & 3.8$\pm$0.5 & & 0.070$\pm$0.020 & 6.5$\pm$1.0 \\
3.1 & \textbf{3.8$\pm$0.7} & 3.7$\pm$0.9 & & 0.056$\pm$0.027 & 5.2$\pm$1.4 
\enddata
\tablenotetext{1}{Bias calculated with $\omega_\mathrm{obs}=b^2\omega_m$ (Equation \ref{eq:linear_bias}), which is the representative measurement in this paper.}
\tablenotetext{2}{Bias calculated with cosmic matter variance, $\sigma_8$ (Equation \ref{eq:bias_sigma})}
\end{deluxetable}

We obtain $b_{\omega} = 4.0 \pm 0.8$ at $z \sim 2.4$ and $3.8 \pm 0.7$ at $z \sim 3.1$. The bias estimates from the two methods are in good agreement.  
We therefore adopt the estimate derived from the ratio with the matter correlation function throughout the rest of this paper as it is defined without reference to a specific scale such as 8\;$h^{-1}$\;Mpc.
Figure~\ref{fig:bias_vs_z} compares the biases of LABs with those in the literature.
LABs show significantly higher biases than LAEs \citep[$b\sim1.5$--2, ][]{Gawiser07, Guaita10, Umeda25, White24, Herrera25} and LBGs \citep[$b\sim2.5$, ][]{Soo06} at $z$ = 2 -- 3, but lower biases than proto-clusters \citep[$b\sim6$,][]{Ramakrishnan25}. 
LAB at $z\sim2.4$ have bias comparable to those of SMGs \citep[$b\sim3.5$,][]{Hickox12, Wilkinson17} and QSOs \citep{Eftekharzadeh15}. On the other hand, LAB at $z\sim3.1$ have smaller bias than SMGs \citep{Wilkinson17} and similar to the lower end of the bias range of QSOs \citep[$b$ = 3.5 -- 7,][]{Shen07, Eftekharzadeh15}.

The only previous bias measurement for LABs is that reported in \cite{Zhang25}, who found $b = 5.3 \pm 0.5$ for 39 LABs at $z\sim2.3$ and $b=6.0\pm1.0$ for 22 bright LABs at the same redshift using the auto-correlation function. 
These measurements are based on far fewer objects than found here, but their higher bias values are still consistent with ours at the $1.4 \sigma$ level.

\subsubsection{Halo Mass of LABs} \label{sec:halo_masses}

\begin{figure}
\centering
\includegraphics[width=0.45\textwidth]{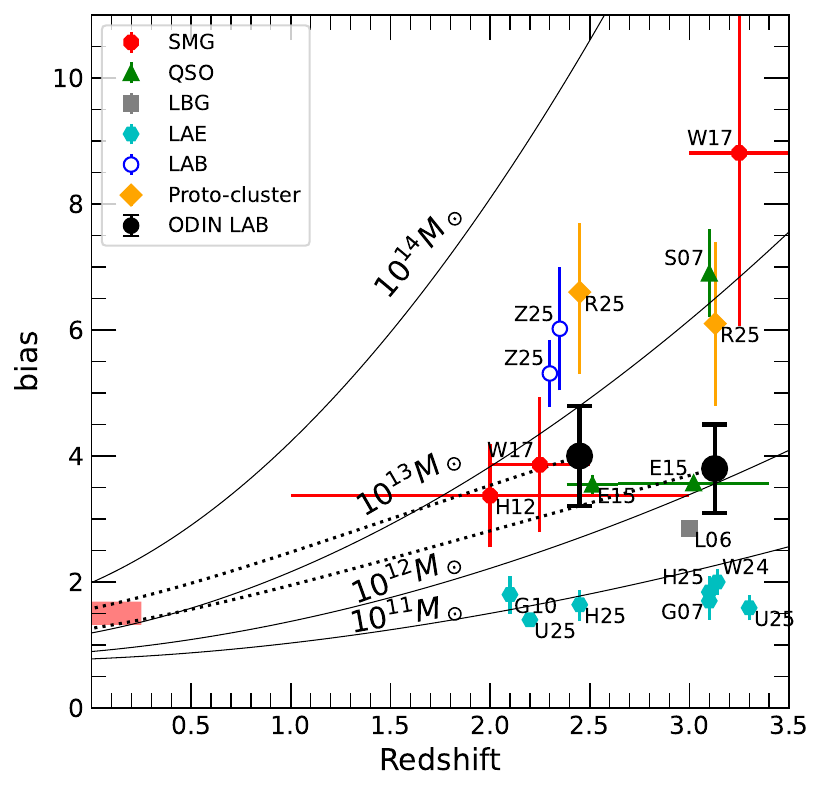}
\caption{Galaxy bias and corresponding halo mass with redshift. The solid lines show the evolution of bias at given dark matter halo mass (Equation~\ref{eq:halo_mass_bias}). 
The orange diamonds represent proto-clusters \citep[R25;][]{Ramakrishnan25} and the red octagons represent SMGs \citep[H12; W17 for][]{Hickox12, Wilkinson17}.
The green triangles are QSOs \citep[S07; E15 for][]{Shen07, Eftekharzadeh15}.
The Grey square shows LBGs \citep[L06;][]{Soo06}.
The cyan hexagons are LAEs obtained from \citep[G07; G10; W24; U25; H25 for][]{Gawiser07, Guaita10, White24, Umeda25, Herrera25}.
The blue open circles represent LABs from \citet[][Z25]{Zhang25}: The lower bias is for their full sample of 39 LABs and the higher bias is for their subset of bright LABs (slightly shifted for visibility).
Our measurements are shown as the large black circles. The dotted lines demonstrate the expected mean growth of dark matter halos \citep{Fakhouri10}.
The red shaded region represent the bias of present day galaxies with (2--4)$L^\star$ \citep{Zehavi11}.
LABs show higher bias than LAEs, but smaller than proto-clusters.
The dark matter halos hosting LABs are expected to evolve into present-day massive halos where low-mass galaxy groups or massive elliptical galaxies reside.
}
\label{fig:bias_vs_z}
\end{figure}

The dark matter halo mass for a given bias is calculated following \cite{Mo_white02}:
\begin{equation} 
\label{eq:halo_mass_bias}
    b(M_{\rm h}) = 1 + \frac{1}{\delta_c}\biggl[\nu^{\prime2}+b\nu^{\prime2(1-c)}-\frac{\nu^{\prime2 c}/\sqrt{a}}{\nu^{\prime2}+b(1-c)(1-c/2)}\biggl], 
\end{equation}
where $\nu^\prime$, rescaled peak-height in the linear density field, is defined as $\sqrt{a}\delta_c/\sigma(M_{\rm h})$, and we adopt the constants $a=0.707, b=0.5, c=0.6$ and $\delta_c=1.69$ \citep{Sheth01}.
The variance in the smoothed density field, $\sigma^2(M_{\rm h})$, is defined using the linear matter power spectrum $P(k, z)$, the Fourier transform of a spherical top-hat filter $W(kR)$, and the Lagrangian halo radius $R$, which depends on the halo mass $M_{\rm h}$:
\begin{eqnarray} \label{eq:variance_density_field}
    \sigma(M_\mathrm{h})^2 &=& \int_0^\infty{\frac{dk}{k} \frac{k^3 P(k,z)}{2 \pi^2} |W(kR)|^2} \\
    W(kR) &=& \frac{3}{(kR)^3}[\sin(kR)-kR\cos(kR)] \nonumber \\
    R &=& \biggl(\frac{3M_\mathrm{h}}{4\pi\rho_0}\biggr)^{1/3}, \nonumber
\end{eqnarray}
where $\rho_0$ is the present mean density of the universe.

For our bias measurements, the halo masses of LABs are $5.1^{+4.6}_{-3.0}\times10^{12}$\;\msun and $1.4^{+1.3}_{-0.8}\times10^{12}\;M_\odot$ for $z\sim2.4$ and $3.1$, respectively (Table~\ref{tab:mass_and_hof}). The uncertainties in halo masses are determined by propagating the uncertainties in the bias measurements using a Monte Carlo approach, assuming that the bias errors are symmetric.

We infer the bias and halo mass of LABs at $z = 0$ ($M^{z=0}_{\rm h}$) by applying the mean mass growth rate of dark matter halos from the Millennium simulation \citep{Fakhouri10}. 
The dark matter halos hosting LABs at $z \sim 2.4$ and 3.1 are expected to evolve into present-day massive halos ($M_{\rm h}\sim10^{13}$\;\msun; dashed lines in Figure~\ref{fig:bias_vs_z}).
Such halos are characteristic of low-mass galaxy groups \citep[][$M_{200}\sim10^{13}~M_\odot$]{Popesso15} or systems hosting 2$L^\star$--4$L^\star$ galaxies (i.e., elliptical galaxies) in the local Universe \citep[red shaded region, $b\sim1.5$;][]{Zehavi11}.

The mass given above corresponds to the top-hat virial mass, in the simplified case in which all objects in a given sample reside in halos of the same mass. 
In the following section, we will calculate the median halo mass of LABs assuming a halo mass function.

\begin{deluxetable*}{c ccc l ccccc}[t]
\centering
\tablecaption{Halo mass and occupation fraction of LABs \label{tab:mass_and_hof}}
\tablewidth{0pt}
\tablehead{
\colhead{} &
\colhead{} &
\multicolumn{2}{c}{Halo mass\tablenotemark{1}} &
\colhead{} &
\multicolumn{5}{c}{Halo Occupation} \\
\cline{3-4}
\cline{6-10} 
\colhead{$z$}  & 
\colhead{$b_{\omega_\mathrm{DM}}$\tablenotemark{2}} &
\colhead{$M_\mathrm{h}$} &
\colhead{$M^{z=0}_{\rm h}$} &
\colhead{} &
\colhead{$n_\mathrm{LAB}$\tablenotemark{3}}  & 
\colhead{$M_\mathrm{h,min}$} &
\colhead{$M_{\rm h,med}$} &
\colhead{$M^{z=0}_{\rm h,med}$} &
\colhead{\hof} \\
\colhead{ }  & 
\colhead{ } & 
\colhead{$10^{12}$ \msun} &
\colhead{$10^{13}$ \msun} &
\colhead{ } & 
\colhead{10$^{-5}$ Mpc$^{-3}$}  & 
\colhead{$10^{12}$ \msun} &
\colhead{$10^{12}$ \msun} &
\colhead{$10^{13}$ \msun} &
\colhead{\%} 
}
\startdata
2.4 & $4.0\pm0.8$ & $5.1^{+4.6}_{-3.0}$ & $4.5^{+4.6}_{-2.8}$ &   & $1.6\pm0.2$ & $2.8^{+3.0}_{-1.8}$ & $4.2^{+3.8}_{-2.5}$ & $3.7^{+5.4}_{-1.9}$ & $11.24^{+39.13}_{-8.15}$ \\
3.1 & $3.8\pm0.7$ & $1.4^{+1.3}_{-0.8}$ & $1.7^{+1.8}_{-1.1}$ &   & $1.7\pm0.2$ & $0.7^{+0.8}_{-0.5}$ & $1.1^{+1.1}_{-0.7}$ & $1.3^{+2.2}_{-0.7}$ & $2.99^{+9.39}_{-2.13}$  \\
\vspace{-0.5cm}

\enddata
\tablenotetext{1}{Halo masses when all LABs reside in halos of the same mass (Section~\ref{sec:halo_masses}).}
\tablenotetext{2}{bias from ACFs with Equation~\ref{eq:linear_bias}.}
\tablenotetext{3}{Observed volume number density of LABs. Uncertainties are determined from Poisson noise of LAB number counts.}
\end{deluxetable*}

\subsubsection{Halo Occupation Fraction and Median Halo Mass}
\label{sec:duty_cycle}
We examine the fraction of dark matter that host LABs and thereby constrain their rarity within the halo population using the HOD framework.
The framework defines the galaxy bias by the probability that a dark matter halo contain $N$ galaxies \citep[e.g.,][]{Berlind02}.

In the simplest case, dark matter halos above a certain minimum mass can host at most one LAB, with a constant halo occupation fraction (\hof). Under this assumption, the halo occupation fraction and the minimum dark matter halo mass of LABs ($M_{\rm h, min}$) can be inferred using the following equations:
\begin{eqnarray}
    \langle b \rangle  &=& \frac{\int_{M_\mathrm{h,min}}^{\infty}{b(M)  n(M) dM} }{\int_{M_\mathrm{h,min}}^{\infty}{  n(M) dM}} \label{eq:bias_for_duty_cycle} \\
    n_\mathrm{LAB} &=& f_\mathrm{LAB} \int_{M_\mathrm{h,min}}^{\infty}{ n(M) dM},\label{eq:ng_for_duty_cycle}
\end{eqnarray}
where $n(M)\,dM$ denotes the halo mass function \citep{Sheth99}, and $\langle b \rangle$ represents the mean bias, which is assumed to be equivalent to the bias factor estimated from Equation~\ref{eq:linear_bias}.
Here, $b(M)$ is the bias factor of dark matter halos as a function of their mass (Equation~\ref{eq:halo_mass_bias}).
Using this formulation, the minimum halo masses of LABs are estimated to be $M_{\mathrm{h,min}}$ = $2.8^{+3.0}_{-1.8}$ $\times$ $10^{12}$ \msun and $7.4^{+8.2}_{-4.6} \times 10^{11}$\msun at $z \sim 2.4$ and 3.1, respectively. 
The corresponding halo occupation fractions (\hof) are $0.11^{+0.39}_{-0.08}$ and $0.03^{+0.09}_{-0.02}$ at $z \sim 2.4$ and 3.1, respectively.
The uncertainties in $M_{\mathrm{h,min}}$ are estimated following the procedure described in Section~\ref{sec:halo_masses}, while the uncertainties in \hof are propagated from those of $M_{\rm h,min}$ and the observed LAB number densities ($n_{\rm LAB}$).

As a representative halo mass scale for LABs, we adopt the median halo mass ($M_{\rm h,med}$), determined from the halo mass distribution down to the minimum halo mass, since median values are more robust than mean values. We obtain  $M_{\rm h,med}$ = $4.2^{+3.8}_{-2.5}$ and $1.1^{+1.1}_{-0.7}\times10^{12}$\;\msun at $z\sim2.4$ and 3.1, respectively. 
The median mass of halos will evolve into halos with a mass of $\sim10^{13}$ \msun by $z=0$.
We summarize the results from the halo occupation analysis in Table~\ref{tab:mass_and_hof}.

\subsection{Cross-correlation Function with LAEs} \label{sec:ccf}

\subsubsection{CCFs of Entire LAB samples} \label{subsec:ccf_all_LAB}
The cross-correlation function (CCF) between different types of galaxies---common (low-mass) galaxies and rare (massive) galaxies---has been used to investigate the clustering properties of rare objects for which the auto-correlation function cannot be reliably measured \citep[e.g., SMGs, QSOs, proto-clusters;][]{Wilkinson17, He18, Ramakrishnan25}.
Although our LAB sample is the largest assembled to date ($\sim$100 objects), the intrinsic rarity of LABs motivates an independent assessment of the clustering measurements. We therefore cross-check our results using the cross-correlation function (CCF) between LABs and the more numerous LAEs.
We use the Landy-Szalay estimator for the CCF, 
\begin{equation}
    \omega_\mathrm{CCF}(\theta) = \frac{D_1\!D_2(\theta)-D_1\!R_2(\theta)-D_2\!R_1(\theta)+R_1\!R_2(\theta)}{R_1\!R_2(\theta)},
    \label{eq:CCF}
\end{equation}
where the subscripts `1' and `2' represent LABs and LAEs, respectively. $D_1D_2$ ($R_1R_2$) are the observed (random) LAB--LAE pair counts at an angular separation of $\theta\pm\Delta\theta/2$, i.e., the same as the ACF definition (Section~\ref{sec:ACF}). $D_1R_2(\theta)$ and $D_2R_1(\theta)$ denote the numbers of observed-random cross pairs for LABs and LAEs, respectively.

We use the ODIN LAE sample from the same field \citep{Firestone24}, and construct the catalog of random LAEs following \citet{Ramakrishnan25} and \citet{Herrera25}. We select 50,000 random sources from the ODIN narrowband photometric catalogs with S/N $>$ 5 to account for small depth variations that could affect LAE selection.
The random catalog for LABs is the same as that used for the ACF (Section~\ref{sec:ACF}).

\begin{figure*}[t]
\centering
\includegraphics[width=\textwidth]{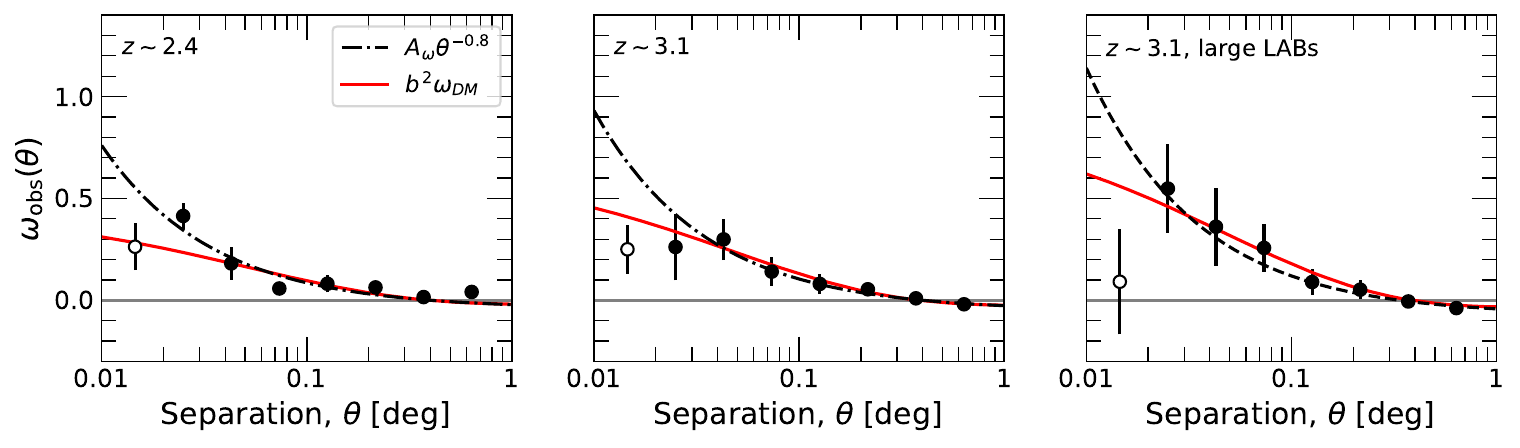}
\caption{
(Left, Middle) Observed CCF between LABs and LAEs at $z \sim 2.4$ and $z \sim 3.1$.
(Right) CCF between large LABs and LAEs at $z \sim 3.1$.
The black dot-dashed lines indicate the best-fit power-law function (Equation~\ref{eq:ACF_fit_func}, $\beta = 0.8$), while the red solid lines show the best-fit results using Equation~\ref{eq:linear_bias}.
Open markers denote data points excluded from the fitting.
}
\label{fig:ccf_fit}
\end{figure*}

Figure~\ref{fig:ccf_fit} shows the observed CCF between LABs and LAEs, together with the best-fit models using Equations~\ref{eq:ACF_fit_func} and~\ref{eq:linear_bias}. We estimate the clustering properties of LABs following the procedure described in Section~\ref{sec:ACF}, adopting the LAE clustering measurements from \cite{White24}.
The cross-correlation function between the two galaxy populations is the geometric mean of their respective ACFs assuming that both populations share the same power-law slope ($\beta$). Therefore, the correlation length and galaxy bias of the LABs are given by
\begin{eqnarray}
    r_{0,\mathrm{LAB}} &=& \frac{r_{0,\mathrm{CCF}}^2}{r_{0,\mathrm{LAE}}} \\
    b_\mathrm{LAB} &=& \frac{b_\mathrm{CCF}^2}{b_\mathrm{LAE}}. 
    \label{eq:geo_mean}
\end{eqnarray}
where the measured correlation lengths ($r_{0,\mathrm{CCF}}$) are $2.7\pm0.4$ and $3.4\pm0.2$~$h^{-1}$\;Mpc for $z\sim2.4$ and $z\sim3.1$, respectively.
Thus, the correlation lengths of LABs inferred from the CCF are $2.5\pm0.8$ and $3.9\pm0.6$ $h^{-1}$\;Mpc at $z\sim2.4$ and $z\sim3.1$, respectively.
The $b_\mathrm{CCF}$ values at $z \sim 2.4$ and $z \sim 3.1$ are measured to be $1.8 \pm 0.3$ and $2.3 \pm 0.1$. 
These values correspond to to LAB biases of $b_{\rm LAB}$ = $1.9 \pm 0.6$ and $2.6 \pm 0.3$, respectively. We note that our results remain consistent when adopting the LAE biases from \cite{Herrera25} rather than those from \cite{White24}.

Halo masses of LABs and their uncertainties are estimated from $b_\mathrm{LAB}$ following Section~\ref{sec:halo_masses}. The characteristic halo masses derived from the CCF analysis are $1.7^{+5.3}_{-1.6} \times 10^{11}$~\msun\ and $2.5^{+1.7}_{-1.1} \times 10^{11}$~\msun\ at $z \sim 2.4$ and $z \sim 3.1$, respectively.
The results of the CCF analysis are summarized in Table~\ref{tab:ccf_result}. 

In summary, we find that the biases and corresponding halo masses derived from the CCF analysis are systematically smaller than those obtained from the ACF analysis.
For the LABs at $z \sim 3.1$, the bias values agree with those from the ACF analysis within $\sim$1.6$\sigma$, whereas the $z \sim 2.4$ sample shows a slightly larger discrepancy of $\sim$2.1$\sigma$. We discuss this further in Section~\ref{sec:ACF_vs_CCF}.

\begin{deluxetable*}{cc ccc l ccccc}[t] 
\centering
\tablecaption{Summary of CCF clustering properties of ODIN LABs \label{tab:ccf_result}}
\tablewidth{0pt}
\tablehead{
\colhead{} &
\colhead{} &
\multicolumn{3}{c}{CCF measurements} &
\colhead{} &
\multicolumn{5}{c}{LAB properties from CCF} \\
\cline{3-5} 
\cline{7-11} 
\colhead{$z$}  & 
\colhead{Samples}  & 
\colhead{$A_\omega$} & 
\colhead{$r_{0,\mathrm{CCF}}$} &
\colhead{$b_\mathrm{CCF}$\tablenotemark{1}} &  
\colhead{} &
\colhead{$r_{0,\mathrm{LAE}}$\tablenotemark{2}} & 
\colhead{$b_\mathrm{LAE}$\tablenotemark{2}} &
\colhead{$r_{0,\mathrm{LAB}}$} & 
\colhead{$b_\mathrm{LAB}$} &
\colhead{$\log{(M_\mathrm{h}}$/\msun)} \\
\colhead{}  &
\colhead{} & 
\colhead{} & 
\colhead{[$h^{-1}$~Mpc]} &
\colhead{} &
\colhead{} &
\colhead{[$h^{-1}$~Mpc]} &
\colhead{} &
\colhead{[$h^{-1}$~Mpc]} &
\colhead{} &
\colhead{} 
}
\startdata
2.4 & All & 0.015$\pm$0.004 & 2.7$\pm$0.4 & 1.8$\pm$0.3 & & 3.0$\pm$0.2 & 1.7$\pm$0.2 & 2.5$\pm$0.8 & 1.9$\pm$0.6 & $11.24^{+1.33}_{-0.39}$ \\
3.1 & All & 0.026$\pm$0.003 & 3.4$\pm$0.2 & 2.3$\pm$0.1 & & 3.0$\pm$0.2 & 2.0$\pm$0.2 & 3.9$\pm$0.6 & 2.6$\pm$0.3 & $11.39^{+0.30}_{-0.20}$ \\ 
3.1 & $A_{\rm iso}>30$ \sqarcsec & 0.038$\pm$0.004 & 4.2$\pm$0.3 & 2.7$\pm$0.2 & & 3.0$\pm$0.2 & 2.0$\pm$0.2 & 5.9$\pm$0.8 & 3.7$\pm$0.6 & $12.09^{+0.36}_{-0.24}$  
\enddata
\tablenotetext{1}{Bias estimated from $\omega_\mathrm{obs}=b^2\omega_m$  (Equation~\ref{eq:linear_bias}).}
\tablenotetext{2}{Values assumed for LAEs \citep{White24}.}
\end{deluxetable*}

\subsubsection{CCFs of Bright/Large LABs} \label{subsec:ccf_large_LAB}
Previous studies of individual bright ($L_{\mathrm{Ly}\alpha} \sim 10^{44}$~\unitcgslum) and large ($\sim$100~\sqarcsec) LABs have reported larger halo masses of $\sim 3.2 \times 10^{13}$~\msun, based on the galaxy--halo mass scaling relation \citep[e.g.,][]{Umehata21, Daddi22}. Furthermore, \cite{Uchimoto12} suggested that the stellar mass of LABs correlates with their \lya luminosities and sizes.
To investigate this possible connection between the \lya properties (sizes and luminosities) and halo mass, we measure the CCF between bright/large LABs and LAEs. 
We note that an auto-correlation measurement for the bright/large LAB subsample is not feasible due to the relatively small sample size. 
In addition, the bright/large LAB sample size at $z \sim 2.4$ is also insufficient for a reliable CCF analysis. 
Therefore, we restrict the CCF analysis to the bright/large LAB subsample at $z \sim 3.1$.

We define the bright/large LAB subsample as those with isophotal sizes ($A_{\rm iso}$) greater than 30~\sqarcsec. This subsample consists of 42 LABs at $z \sim 3.1$ with $L_{\mathrm{Ly}\alpha} > 1.5\times10^{43}$~\unitcgslum.
For the bright/large sample, we measure the correlation length, bias, and halo mass using the same method described in Section~\ref{subsec:ccf_all_LAB}.
Figure~\ref{fig:ccf_fit} (right) shows the best-fit result of the CCF (red line), with a bias of $b_\mathrm{CCF} = 2.7 \pm 0.2$, yielding $b_\mathrm{LAB} = 3.7 \pm 0.6$ and an inferred characteristic halo mass of $1.2^{+1.0}_{-0.7} \times 10^{12}$~\msun. 
The CCF correlation length is $r_{0,\mathrm{CCF}} = 4.2 \pm 0.3$~$h^{-1}$\;Mpc, which implies $r_{0,\mathrm{LAB}} = 5.9 \pm 0.8$~$h^{-1}$\;Mpc.
Although our measurements are still limited by small-number statistics, the results suggest that the bright/large LABs reside in halos that are $\sim$5 times more massive ($1.2^{+1.0}_{-0.7} \times 10^{12}$~\msun) than those of the full sample ($2.5^{+1.7}_{-1.1} \times 10^{11}$~\msun). These findings are consistent with \cite{Zhang25}, who divided their $N = 39$ sample into a bright/large subsample ($N = 22$) and measured the ACFs separately.

\section{Discussion \label{sec:discussion}}

\subsection{Discrepancy Between Bias Estimates from Auto- and Cross-correlation function} 
\label{sec:ACF_vs_CCF}

In Section~\ref{sec:ccf}, the biases of LABs derived from the CCF measurements are lower at both $z \sim 2.4$ and 3.1 than those obtained from the ACF measurements, although the significance of this discrepancy is only 2.1$\sigma$ and 1.6$\sigma$, respectively. Here, we discuss possible reasons for the difference.

First, the discrepancies could simply be attributed to the small sample sizes and resulting statistical fluctuations.
Figure~\ref{fig:acf_fit} shows significant scatter in the ACFs at both $z \sim 2.4$ and $z \sim 3.1$.
In the case of the CCFs, the $z \sim 3.1$ LABs show good agreement between all data points and the model fits. However, the $z \sim 2.4$ LABs still exhibit substantial scatter around the models.
Therefore, the $z \sim 3.1$ measurements are likely reliable within the 1.6$\sigma$ level, whereas the $z \sim 2.4$ results require larger samples for confirmation.
We expect that the auto- and cross-correlation functions derived from the full ODIN LAB sample will reduce these scatters and mitigate the current discrepancies.

Second, the construction of the random catalog for LAB could also contribute to the discrepancy. Although we use the inverse variance maps of the narrowband images to account for surface-brightness fluctuations, the actual detection efficiency and number density of LABs may be affected by additional factors, such as the surface-brightness profiles and morphologies of the LABs. With larger ODIN samples in the future, reliable LAB spatial profile will enable more realistic recovery tests on the \lya images. This will allow us to construct a more robust random catalog that accounts for the actual detectability of LABs at a given detection threshold and, in turn, yield more reliable auto-correlation function measurements.

On the other hand, if the discrepancies are intrinsic, this would indicate that LABs cluster differently from the predictions of the simple linear bias model \citep[$\delta_g = b\,\delta_m$, where $\delta_g$ and $\delta_m$ are the galaxy and matter overdensities;][]{Kaiser84, Bardeen86}. Instead, their clustering might be influenced by non-linear ($\delta_g = b\,\delta_m + \delta_m^2\,b^2/2$) and/or stochastic bias that reflects the influence of small-scale perturbations on the formation of galaxies \citep[e.g.,][]{Fry93,  Dekel99, Taruya99, Somerville01}. 
The non-linear and/or stochastic bias could reduce the cross-correlation bias below the geometric mean between the two ACFs as shown in the previous observation \citep[e.g.,][]{Wang07, Zehavi11}.

We speculate that the complex galaxy bias may be linked to the spatial distribution of LABs.
\citet{Ramakrishnan23} found that most LABs ($\sim$70\%) are strongly associated with cosmic filaments, typically residing at projected distances within $<$2 Mpc.
They further demonstrated that the apparent association between LABs and protoclusters is a by-product of the converging filamentary structures where protoclusters commonly form.

The robust auto- and cross-correlation analyses using the final ODIN LAB samples ($N$ $\sim$ 1,000) will clarify whether the observed discrepancies are genuine and whether non-linear clustering processes must be taken into account. In addition, examining the spatial distributions of LAEs, LABs, proto-clusters, and filaments will yield valuable insights into the nature of large-scale structure and aid in the interpretation of the cross-correlation function.

\subsection{Comparison with Halo Mass Estimates from Previous Studies} 
\label{sec:comparison}

In this work, we estimate the dark matter halo masses, their redshift evolution, and halo occupation fractions of LABs using the largest homogeneous sample currently available. While \cite{Zhang25} measured the bias of LABs based on a limited sample, their study did not derive halo mass estimates or explore the implications for halo occupation. By deriving halo properties, our study provides a more quantitative characterization of the host halo properties of LABs and enables clearer comparison with results obtained using other methods and samples.

We compare the dark matter halo properties of LABs with estimates from previous studies. 
Due to their rarity, prior to the clustering analyses (this work and \citealt{Zhang25}), the halo masses of LABs had been inferred either from simple clustering analyses based on field-to-field variations \citep{Yang10} or from the stellar-to-halo mass relation (SHMR) \citep{Umehata21}.
Here, we focus on two well-studied LAB samples, each employing one of these methods: 
the LABs discovered in the Chandra Deep Field South \citep[CDFS LABs;][]{Yang10} at $z\sim2.3$ and 
the SSA22 LABs \citep{Matsuda04} at $z\sim3.1$, both of which span a wide range of \lya luminosities and sizes.

First, we compare our results with the minimum halo mass and halo occupation fraction of LABs derived by \citet{Yang10}. Their estimation method is essentially the same as our HOD framework in that it compares the observed LAB number density and field-to-field variance with predictions from an $N$-body simulation \citep[ABACUS;][]{Metchnik09}. Note that their `detectability fraction' corresponds to the halo occupation fraction (\hof) in our framework.
They counted halos within small cells in the simulation box (the counts-in-cells, or C-in-C method) and determined the minimum halo mass required to reproduce the observed LAB number density for a given \hof. The minimum halo mass and \hof were constrained by finding the values that best reproduced the observed field-to-field variance ($\sigma_v = 1.5$), based on the observed distribution of $N_{\rm LAB} = (6, 0, 0, 0)$ across four survey fields. They concluded that \hof\ $> 0.5$ and $M_{\rm min}$ $>$ $1.2\times10^{13}$ \msun are required to achieve a model variance of $\sigma_v (\mathrm{fit}) = 0.93$, comparable to the observed value. These estimates are somewhat larger than our inferred halo masses and \hof.

Thanks to the greatly improved survey statistics of ODIN, we are now able to revisit the counts-in-cells analysis. To enable a direct comparison with their estimates, we measure the field-to-field variance of LABs at $z \sim 2.4$, which closely matches the redshift of the \citet{Yang10} sample.
First, we conduct a recovery test following \citet{Yang10} and \citet{Moon26a} to assess how many CDFS LABs would be recovered at the ODIN survey depth. We confirm that the ODIN survey depth at $z$ = 2.4 is comparable to that of the six bright CDFS LABs. 
Next, we divide the 9~\sqdeg COSMOS field into 72 subregions, each having the same survey volume ($1.2 \times 10^{5}$\;Mpc$^3$) as that of \cite{Yang10}, corresponding to a $22.1 \times 22.1$~\sqarcmin region.
We compute the surface-density variation of LABs following \citet{Ouchi08} and \citet{Yang10}:
\begin{equation}
    \sigma^2_{v}=\frac{\langle n^2 \rangle - \langle n \rangle^2}{\langle n \rangle^2} - \frac{1}{\langle n \rangle},
\end{equation}
where $n$ is the effective surface number density within each subregion.
The measured field-to-field variation is $\sigma_{v,\mathrm{ODIN}} = 0.68 \pm 0.12$.
The uncertainty is estimated using jackknife resampling, as described in Section~\ref{sec:ACF}.

The new field-to-field variance measurement corresponds well to the lower end of the estimates in \citet[][Table~7]{Yang10}, where $\sigma_v$, \hof, and $M_{\rm min}$ are 0.76, 0.125, and $5.3 \times 10^{12}$\,\msun, respectively.
These quantities are consistent with our results within the $1\sigma$ range: 
($\sigma_{v,\mathrm{ODIN}}$, \hof, $M_{\rm h,min}$) = 
($0.68 \pm 0.12$, $0.11^{+0.39}_{-0.08}$, $2.8^{+3.0}_{-1.8} \times 10^{12}\,M_\odot$).
Therefore, we conclude that with the improved LAB statistics, the original \citet{Yang10} estimates can be revised slightly downward, bringing them into better agreement with our clustering analysis.

Second, we compare our results with halo masses inferred from the stellar masses derived for the SSA22 LAB sample at $z\sim3.1$ \citep{Matsuda04}. The SSA22 LAB sample serves as an ideal comparison dataset because the properties of its members have been extensively characterized \citep[e.g.,][]{Uchimoto12, Kubo16, Kato18}.

For a statistical comparison, we estimate the halo masses of the SSA22 LABs following the same methodology as in \citet{Umehata21} and references therein.
In this approach, the K-band-derived stellar masses of the four most massive member galaxies within SSA22-LAB1 are summed and then converted into a halo mass using the known stellar-to-halo mass relation \citep{Durkalec15}.
\citet{Uchimoto12} conducted a deep near-infrared (NIR) survey and confirmed that 15 out of 20 LABs host one or more $K$-band–selected galaxies using photometric and spectroscopic redshifts. The stellar masses for these galaxies were derived through SED fitting using multi-band photometry ($u^*\,B\,V\,R\,i^\prime\,z^\prime\,J\,H\,K$). These 15 LABs provide a dataset for the comparison, spanning a wide range in \lya\ luminosities ($L_{{\rm Ly}\alpha}>0.7\times10^{43}$ \unitcgslum) and spatial extent ($> 17$\,\sqarcsec).
We estimate the halo masses of the 15 SSA22 LABs by adopting a SHMR of $1.32^{+0.98}_{-0.57}\times10^{-2}$ from \citet{Durkalec15}. The uncertainties in the SHMR and stellar mass estimates are propagated into the resulting halo mass estimates. The derived physical properties of the SSA22 LABs are summarized in Table~\ref{tab:SSA22}.

We find that the representative halo masses estimated from stellar mass measurements of the SSA22 LABs are comparable to those derived from our ACF analysis.
The median halo mass of SSA22 LABs, $M_{\rm h,med}$ = $3.9^{+2.8}_{-1.7}\times10^{12}$\,\msun, is in a good agreement with the ODIN LABs at $z\sim3.1$ ($2.1^{+2.0}_{-1.2}\times10^{12}$\,\msun).
Except for SSA22-LAB8 ($M_h=3.2\times10^{11}$~\msun), which may be part of SSA22-LAB1, the individual halo mass estimates are larger than the minimum halo mass at $z\sim3.1$ derived in this work ($M_\mathrm{h,min} = 9.4\times10^{11}$~\msun). 
If we restrict the SSA22 sample to bright/large LABs detectable at the ODIN survey depth, five out of the 15 $K$-band–detected LABs are recovered, yielding a higher median halo mass of $7.0^{+7.6}_{-4.4}\times10^{12}$\;\msun, which is still consistent with our ACF measurements within 1.3$\sigma$.

\begin{deluxetable}{lccrr}[t]
\centering
\tablecaption{SSA22 LAB properties \label{tab:SSA22}}
\tablewidth{0pt}
\tablehead{
\colhead{ID} & 
\colhead{$L_{\rm{Ly}\alpha}$} &
\colhead{Size} &
\colhead{$\log(M_\star/M_\odot$)} & 
\colhead{$\log(M_{\rm h}/M_\odot$)} \\
\colhead{} & 
\colhead{$10^{43}$~erg~s$^{-1}$} &
\colhead{arcsec$^2$} &
\colhead{} & 
\colhead{} \\
\colhead{(1)} & 
\colhead{(2)} & 
\colhead{(3)} &  
\colhead{(4)} & 
\colhead{(5)} 
}
\startdata
 LAB01 &   11.0 &  222 & $11.45^{+0.15}_{-0.15}$ & $13.33^{+0.36}_{-0.24}$\\
 LAB02 &    8.5 &  152 & $11.44^{+0.91}_{-0.19}$ & $13.32^{+0.97}_{-0.27}$\\
 LAB03 &    5.8 &   78 & $10.97^{+0.34}_{-0.20}$ & $12.85^{+0.47}_{-0.27}$\\
 LAB05 &    1.7 &   55 & $10.86^{+0.15}_{-0.15}$ & $12.74^{+0.36}_{-0.24}$\\
 LAB07 &    1.5 &   40 & $10.34^{+0.74}_{-0.23}$ & $12.22^{+0.81}_{-0.30}$\\
 LAB08 &    1.7 &   39 &  $9.62^{+2.40}_{-0.42}$ & $11.50^{+2.42}_{-0.46}$\\
 LAB11 &    0.9 &   30 & $10.20^{+1.08}_{-0.24}$ & $12.08^{+1.13}_{-0.30}$\\
 LAB12 &    0.9 &   29 & $10.55^{+0.33}_{-0.29}$ & $12.42^{+0.46}_{-0.34}$\\
 LAB14 &    1.2 &   27 & $11.23^{+0.13}_{-0.25}$ & $13.11^{+0.35}_{-0.32}$\\
 LAB16 &    1.0 &   25 & $10.43^{+0.66}_{-0.25}$ & $12.31^{+0.73}_{-0.32}$\\
 LAB20 &    0.6 &   21 & $10.19^{+0.39}_{-0.36}$ & $12.07^{+0.50}_{-0.41}$\\
 LAB24 &    0.8 &   19 & $10.45^{+0.42}_{-0.26}$ & $12.33^{+0.53}_{-0.32}$\\
 LAB27 &    0.7 &   18 & $10.86^{+0.26}_{-0.21}$ & $12.74^{+0.42}_{-0.28}$\\
 LAB30 &    0.9 &   17 & $10.49^{+0.52}_{-0.23}$ & $12.37^{+0.61}_{-0.30}$\\
 LAB31 &    1.1 &   17 & $10.41^{+0.33}_{-0.27}$ & $12.29^{+0.46}_{-0.33}$\\
\hline
Median &    1.0 &   22 & $10.49^{+0.52}_{-0.23}$ &   $12.37^{+0.61}_{-0.30}$\\
\vspace{-0.5cm}

\enddata
\tablecomments{(1)--(3) LAB ID, \lya luminosity and size of LABs published in \citet{Matsuda04} (4) Total stellar mass within LABs from \citet{Uchimoto12} (5) Dark matter halo mass derived with SHMR.}
\end{deluxetable}

The agreement between our results and previous estimates based on the C-in-C method and SHMR indicates that LABs are hosted by massive dark matter halos ($M_\mathrm{h}\gtrsim 10^{12}$\,\msun), thereby supporting the robustness of our halo mass estimates.

\subsection{Evolution of halo occupation fraction} \label{sec:hof_evolve}
The results of HOD framework (Table~\ref{tab:mass_and_hof}) suggest an evolution of halo mass and halo occupation fration (\hof) over cosmic time.
LABs at $z\sim2.4$ reside in halos that are approximately four times more massive and exhibit $\sim$4 times higher \hof than those at $z\sim3.1$.
The higher \hof at $z\sim2.4$ arises from the similar LAB number density but a smaller halo number density ($n_\mathrm{halo}=\int n(M)dM$) compared to $z\sim3.1$. 
This implies that LABs are more frequently observable in massive halos at $z\sim2.4$, despite the smaller halo number density.

To further investigate the evolution of halo occupation in terms of powering sources, we estimate the lifetime of LABs following \cite{Herrera25}. Since continuous \lya photon production is required to maintain the LAB phase, the lifetime of the energy sources is expected to be comparable to that of LABs.
The lifetime of LAB phases are estimated to be $296_{-239}^{+631}$~Myr at $z\sim2.4$ and $61_{-49}^{+116}$~Myr at $z\sim3.1$. These values are larger than the lifetimes of potential energy sources: SMGs with $\sim$100~Myr at $z\sim2$ \citep{Hickox12}; QSOs with $\sim$3~Myr at $z\sim2.5$ \citep{Eftekharzadeh15} and $\sim$45~Myr at $2.9<z<3.5$ \citep{Shen07}.
The longer lifetime of LABs may suggest that the LAB phase is more complex than being powered by a single source or mechanism, and could be influenced by multiple mechanisms such as cooling flows \citep{Rosdahl12} or galaxy mergers \citep{Yajima13}. However, drawing firm conclusions remains difficult due to the large uncertainties in the inferred LAB lifetimes.
To further understand the LAB lifetime, halo occupation, and powering mechanisms, measurements of the halo occupation fraction with smaller uncertainties from larger samples are required. 


\section{Summary} \label{sec:summary}
Using the largest LAB sample to date, we investigate the clustering and dark matter halo properties of LABs through measurements of the angular two-point auto correlation function for 103 LABs at $z\sim2.4$ and 112 LABs at $z\sim3.1$ discovered in the extended COSMOS field. 
\begin{itemize}
    \item 
    We obtain bias factors of the LABs: $b=4.0\pm0.8$ and $3.8\pm0.7$ at $z\sim2.4$ and 3.1, respectively. Assuming a power-law angular correlation function with a fixed slope of $\beta=0.8$, we derive correlation lengths of $r_0$ = $6.5\pm1.0$ and $5.2\pm1.4$~$h^{-1}$\;Mpc for LABs at $z\sim2.4$ and $z\sim3.1$, respectively. 

    \item The inferred median halo masses of LABs are $5.1^{+4.6}_{-3.0}\times10^{12}$\;\msun and $1.4^{+1.3}_{-0.8}\times10^{12}$\;\msun at $z\sim2.4$ and 3.1, respectively. If we adopt a mean halo mass growth rate from simulations, the halos hosting LABs are likely to evolve into massive halos by $z=0$ ($M_\mathrm{halo}\sim10^{13}$\;\msun), corresponding to the halo mass scale of massive elliptical galaxies and/or low-mass galaxy groups. 

    \item By assuming a simplified HOD, we estimate the minimum halo masses of LABs to be $2.8^{+3.0}_{-1.8}\times10^{12}$\;\msun and $7.4^{+8.2}_{-4.6}\times10^{11}$\;\msun for $z\sim2.4$ and 3.1, respectively. The corresponding halo occupation fractions \hof are $\sim$0.11$^{+0.39}_{-0.08}$ and $0.03^{+0.09}_{-0.02}$ for $z\sim2.4$ and $z\sim3.1$, respectively.

    \item We compare the inferred halo properties with those in the literature, finding good overall agreement.
    In particular, LABs at $z \sim 2.4$ exhibit field-to-field variations, halo occupation fractions, and minimum halo masses comparable to those reported by \citet{Yang10} based on the counts-in-cells method. At $z \sim 3.1$, the inferred dark matter halo masses of LABs are consistent with those of the SSA22 LABs, whose halo masses were derived using the stellar-to-halo mass relation. These results support that LABs reside in more massive halos ($M_{\rm h} \gtrsim 10^{12}$ \msun) than typical LAEs and LBGs.

    \item Measurements from the cross-correlation function between LABs and LAEs yield smaller bias factors than those derived from the auto-correlation function, with $b = 1.9 \pm 0.6$ and $2.6 \pm 0.3$ at $z \sim 2.4$ and 3.1, respectively. The observed differences can be attributed to limited sample sizes, random catalogs, or the presence of non-linear and stochastic bias effects, possibly related to the tendency of LABs to reside preferentially near filaments of the large-scale structure \citep{Ramakrishnan23}. A robust joint analysis of auto- and cross-correlation functions using the full ODIN samples will be essential to draw firm conclusions.

\end{itemize}

\begin{acknowledgments}
Based on observations at Cerro Tololo Inter-American Observatory, NSF’s NOIRLab (Prop. ID 2020B-0201; PI: K.-S. Lee), which is managed by the Association of Universities for Research in Astronomy under a cooperative agreement with the National Science Foundation.
B.M. and Y.Y. are supported by the Basic Science Research Program through the National Research Foundation of Korea funded by the Ministry of Science, ICT \& Future Planning (2019R1A2C4069803), and UST Young Scientist+ Research Program 2022 through the University of Science and Technology (2022YS45).
E.G., N.F., and D.H. acknowledge support from NSF grant AST-2206222. 
N.F. acknowledges support from the NSF Graduate Research Fellowship Program under Grant No. DGE-2233066. 
K.S.L. acknowledges financial support from the National Science Foundation under grant Nos. AST-2206705, AST-2408359, and from the Ross-Lynn Purdue Research Foundations.
The Institute for Gravitation and the Cosmos is supported by the Eberly College of Science and the Office of the Senior Vice President for Research at the Pennsylvania State University. R.C. and C.G. acknowledge financial support from the National Science Foundation grant AST-2408358.
S.L. acknowledges support from the National Research Foundation of Korea (NRF) grants (2021M3F7A1084525 and RS-2025-00573214) funded by the Korea government (MSIT).
H.S. was supported by the National Research Foundation of Korea (NRF) grant funded by the Korea government (MSIT) (No. RS-2025-25442707). 
L.G. gratefully acknowledges support from the FONDECYT regular project number 1230591, the ANID BASAL project FB210003, ANID-MILENIO-NCN2024\_112
H.S.H. acknowledges the support of the National Research Foundation of Korea (NRF) grant funded by the Korea government (MSIT), NRF-2021R1A2C1094577, and Hyunsong Educational \& Cultural Foundation.

\facilities{Blanco (DECam)}

\end{acknowledgments}


\bibliography{master.LAB}{}
\bibliographystyle{aasjournalv7}




\end{document}